\documentclass[12pt,preprint]{aastex}

\newcommand{\myemail}{jbp@cmu.edu}

\begin{document}

\title{Retroactive Coherence}

\author{J. B. Peterson\altaffilmark{1}}
\affil{Department of Physics, Carnegie Mellon University, Pittsburgh, PA 15213}

\altaffiltext{1}{email:\myemail}

\begin{abstract}
In gravitational microlensing, interference visibility is established
when the spectral resolution of the observation
is less than the reciprocal of the time delay difference. Using high resolution
spectroscopy fringes will be visible at large path differences,
even when the source emission is incoherent.  This is possible because
coherence can be established after interference has already occurred.
Results of a 
laboratory test are presented which demonstrate 
this retroactive establishment of coherence. 
\end{abstract}


\section{Introduction}

Interference can occur when there are two paths for light propagation 
from a source to an observer. 
In gravitational microlensing (\cite{a00}, \cite{w01}, \cite{d99}) 
wavetrains from a point on the source star
travel on two paths,  
and the wave amplitudes are combined before the light is
detected. The lensing geometry has the potential to produce interference,
and applications of gravitational lensing interference have been
considered for wavelength regimes from radio waves to gamma rays
(\cite{man81}, \cite{ss85}, \cite{dw88}, \cite{pf91}, \cite{g92}, \cite{spg93},
 \cite{ug95}).
As many of these authors note, interference visibility will
be diminished if the difference of the length of the two paths
exceeds the coherence length of the light. 

The light sources in gravitational microlensing are incoherent.
In many microlensing examples the light sources are stars and stellar 
photospheres emit primarily blackbody radiation. 
The temporal coherence length of this light is $L_{coh}= c~t_{coh} = c / \Delta \nu$,
where $c$ is the speed of light, $t_{coh}$ 
is the coherence time and $\Delta \nu$ is the half-power bandwidth 
of the source emission spectrum. 
Because of the large bandwidth of the 
blackbody spectrum, the coherence length $L_{coh}$ is about equal to
the peak wavelength. Blackbody emission is often labeled ``incoherent''.

Blackbody emission can be modeled as a continuous 
stream of short wavepackets. 
Figure 1 shows wavepackets of length $L_{coh}$ traveling on two
paths from source to observer around a lensing mass. 
Because the wavepackets do not overlap at arrival, 
the short coherence length of the light would seem to 
indicate that gravitational lensing
interference is undetectable except in cases where the path length
difference amounts to just a few wavelengths. However, this is not 
correct, as the experiment described below demonstrates.

Coherence can be established retroactively. {\it After}
incoherent light has passed through the interference apparatus 
(in the lab or in the Universe), it can be filtered to a narrow bandwidth and
that filtered light will have a new, longer coherence length.  This allows
gravitational microlensing interference to be observed for large path 
length differences.

In this Letter I present results of a laboratory experiment, carried out 
using a Michelson interferometer, which demonstrates retroactive coherence.  

\section{Two Path Coherence}

An analysis of partial coherence for two path geometries is presented in 
\cite{bw80}. I adapt their analysis to the present purpose.

To describe the degree of interference seen in the patterns formed by his
interferometers
Michelson defined a quantity called the
interference visibility
$V(\Delta \ell)  = (I_{max}-I_{min})/(I_{max} + I_{min})$.  
Here the maximum and minimum of the intensity $I_{max}$
and $I_{min}$ 
are taken as the
path length difference $\Delta \ell$
varies. $V$ ranges from 0 to 1, and generally has a maximum
at zero path difference. 

For two-path gravitational microlensing, or for light that passes through a 
Michelson interferometer, the detected intensity is 

\begin{equation}
I(\Delta \ell)= <|\sqrt{I_1} f(t) + \sqrt{I_2} f(t+\Delta \ell/c)|^2 >_T
\end{equation}

where $I_1$ and $I_2$ are the intensities on the two paths, $f(t)$
is
a complex function describing the time variation of the arriving field strength
on the shorter path, and
$f(t + \Delta \ell/c)$ is the same function shifted in time by the path delay
difference.
$f$ is normalized, i.e. $<f^*(t)f(t)>=1$, and assumed to be stationary.
$T$ is the duration of the observation, the time over which the
intensity is averaged, and is taken to be much longer than
$t_{coh}$ and $\Delta \ell/c$.

The total intensity is then

\begin{equation}
I(\Delta \ell) = I_1  + I_2 + 2 \sqrt{I_1 I_2} Re (<f^*(t)f(t+ \Delta \ell/c)  >_T)
\end{equation}

The last term describes the interference is also called the 
real part of the complex degree of coherence 
$\gamma(\Delta \ell) = <f(t)^*f(t+\Delta \ell/c)>$.

For an optical frequency band centered at $\nu_c$ the 
magnitude of the degree of coherence
is related to the interference visibility by 
$Re (\gamma(\Delta \ell)) = V(\Delta \ell) cos (2 \pi \nu_c \Delta \ell /c)$.

The autocorrelation of the wave amplitude $\gamma$ is the 
fourier transform of the power spectral density $P(\nu)$ 
of the light and the coherence length is the width of the function 
$|\gamma_{12}(\Delta \ell)|$. 
This means that decreasing the bandwidth of
the observation, i.e. forcing $P(\nu)$ to zero outside a narrow range of
frequencies, increases the coherence length and makes interference visible at
large path length differences. The experiment described below shows that this 
is true even when the bandwidth restriction occurs after the light has 
passed through the interference apparatus.

\section{Laboratory Analog of Lensing Interference}

I set up a laboratory analog of gravitational lensing interference using a
Michelson interferometer, as shown in figure 2. In the experiment
light from a 
laboratory thermal source 
arrives at a partially 
reflecting beam splitter which divides the light among two paths. 
These paths are the two orthogonal arms of the interferometer. 
At the ends of the arms mirrors return the light to the 
beam splitter with a path 
length difference that can be adjusted.  
The beam splitter recombines the light before it passes through 
the exit aperture.

Filtering the light restores interference visibility.
In gravitational lensing the path length varies with time because of
relative motion among the source, lensing masses, and the observer.
In the Michelson interferometer, the path lengths 
are stable
but the interferometer can be adjusted to give various path lengths through
various parts of the instrument.
I tilted one mirror slightly so zero path length difference
occurs only at a line dividing the pattern vertically.  
The pattern is also divided horizontally.
The top half of the exit aperture of the interferometer is unblocked and here 
the full bandwidth of the thermal source contributes to the 
interference pattern.
The bottom half of the exit aperture is covered by a narrow-band 
($486 nm \pm 15 nm$) filter. In figure 3 the results of the experiment are 
displayed. For white light the visibility falls almost to zero within a few
fringes. For filtered light the visibility remains near one for 
the entire width of the pattern.

Considering the frequency domain it is easy to see
how coherence is restored through bandwidth restriction: the white light 
interference pattern 
is composed of many narrow-band patterns overlaid. The spacing of
the fringes is proportional to wavelength.
For all wavelengths, the patterns are aligned in the center of the image,
at zero path difference. 
The central few fringes have high visibility but, away from the
center of the image, the pattern develops colored stripes, as red and
blue fringe patterns separate in phase.  
At the edge of the pattern the visibility has
fallen nearly to zero because of cancellation, 
among the various wavelengths, of constructive and destructive 
interference.  By selecting a narrow range of frequencies
cancellation is avoided and the interference becomes 
visible across the entire field. This is shown 
in the right panel of figure 3.

In the time domain a puzzle remains: how can short coherence length
wave packets, 
which don't overlap as they exit the interferometer, create visible
interference at large path length differences?  
The answer is that the filter stores multiply delayed
copies of the wave train incident upon it. 
Interference occurs among these copies.  Light storage in a filter is 
illustrated in figure 4.

\section{Discussion}

In the experiment described here a single narrow filter is used to
restrict the passband of the observation. Most of the optical power 
from the source is
blocked by this filter. In a gravitational microlensing observation a
spectrophotometer could be used instead of a filter. This would
divide the spectrum into many
narrow passbands, each acting as a narrow filter.
This allows all the optical power to be used in the observation.
For a spectrophotometer the coherence time $t_{coh}$ is the reciprocal of the
frequency resolution.

Although temporal coherence can be increased by using high spectral 
resolution, limited {\it lateral} spatial coherence remains a factor reducing
lensing interference visibility for most geometries studied so far. 
As described in the references above, the path length 
difference varies from point to point across
the disk of a source star. The total variation is generally many wavelengths,
and this substantially reduces interference visibility.  

Occasionally, microlensing events include caustic crossings.  
At the first crossing two new, high magnification paths appear. These
paths have the same length at first so
the size of the laterally coherent region (\cite{jp95}, \cite{gg97} 
) can cover a much larger part of the source disk than in the more common
Schwarzschild case.

A Michelson interferometer is useful, not just as a laboratory analog
of gravitational lensing interference, but also as a tool for observation
of the effect. Microlensed light, collected with a telescope, can be
passed through a Michelson interferometer before detection. Then,
if the interferometer path length difference matches the lensing
path length difference, the two arriving wavepackets seen in figure 1
can be realigned, and bandwidth reduction of visibility can be avoided.
For slight mismatches of the two path length difference the 
required bandwidth for full visibility is 
$\Delta \nu < c/ (\Delta \ell_{interferometer} - \Delta \ell_{lensing})$.

Even though a complex representation of the wave amplitude is used
in this discussion of interference, and even 
though the uncertainty principle $\Delta t_{coh} ~ \Delta \nu \sim 1$ 
applies to the coherence time, 
the effects described here are classical rather than quantum 
mechanical. No use is made, in this analysis, of photon quanta.

The use of high resolution spectroscopy establishes coherence
retroactively, allowing gravitational microlensing interference 
to be visible for path length differences many times larger than the
wavelength of the light.

\clearpage

\begin{figure}
\plotone{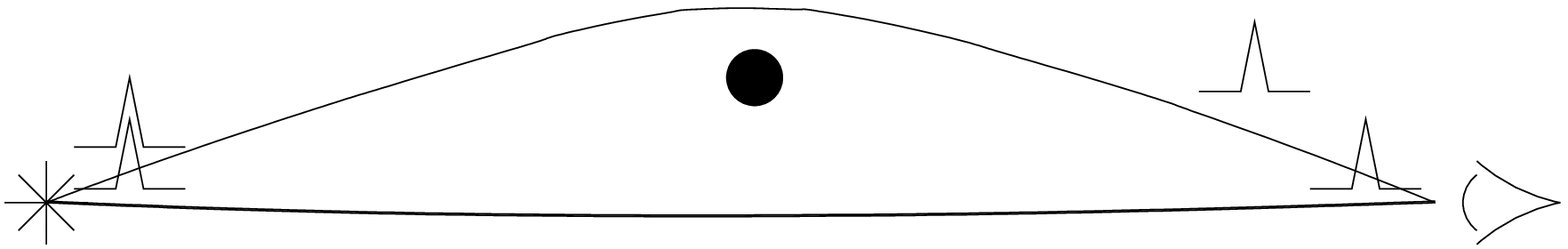}
\caption{{\bf Gravitational microlensing time delay.}
In gravitational microlensing light travels on two
paths, and is recombined before detection.
The spike waveforms shown represent the short coherence length of a
wavepacket from a blackbody source.  If the lengths of the two paths differ
by more than the coherence length, and the detector is sensitive to the 
full bandwidth of the arriving light, there will be strong attenuation of the
interference visibility.
\label{lens}}
\end{figure}

\clearpage 

\begin{figure}
\plotone{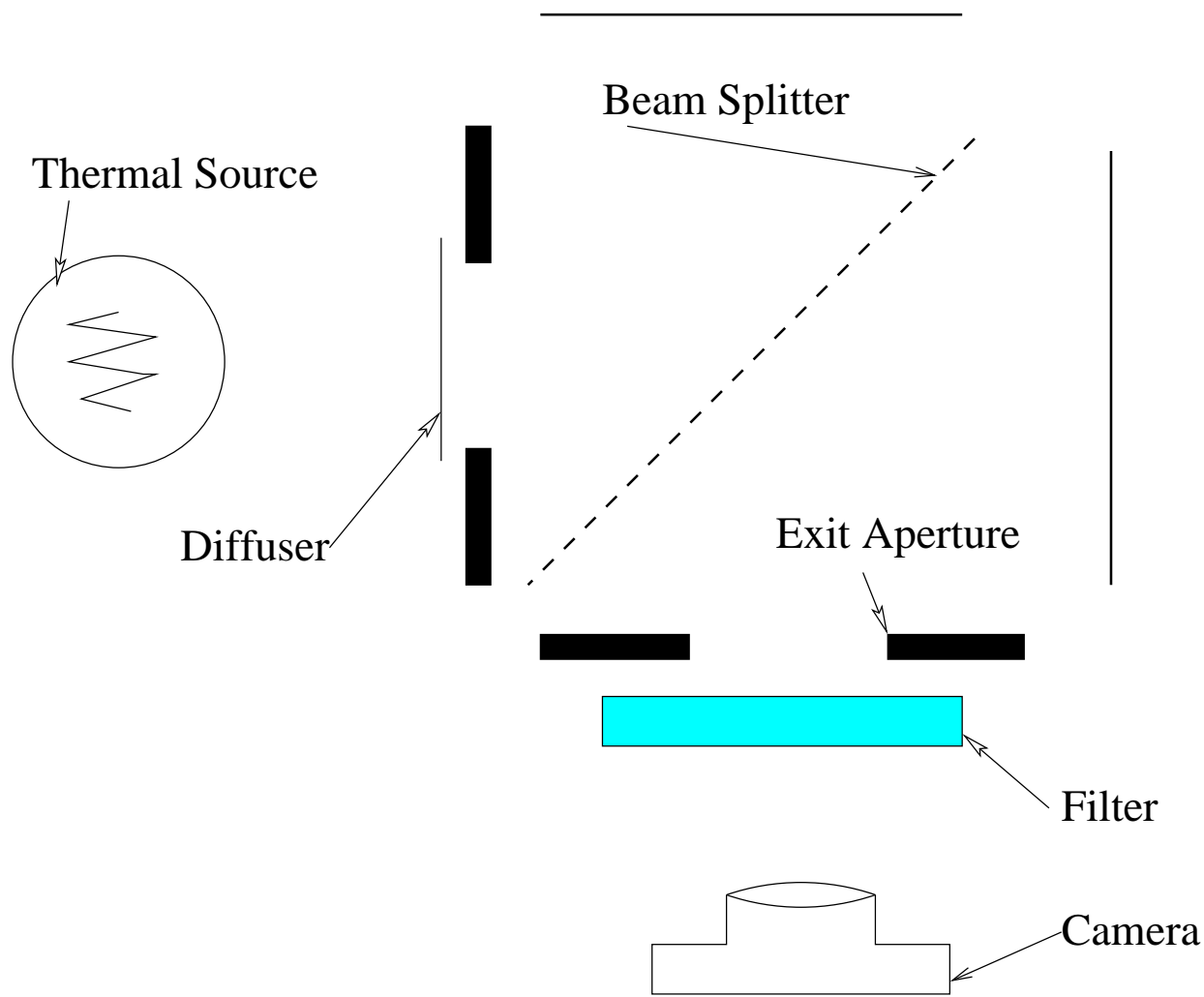}
\caption{{\bf Michelson interferometer used to demonstrate retroactive 
coherence.}
A thermal light source illuminates a ground glass diffuser.
Light from each point on the diffuser 
is divided by the beam splitter, reflects off the
flat mirrors and returns to the beam splitter.  Light amplitudes are
recombined at the beam splitter and the composite waves pass out the
exit aperture.  A filter processes the wave, after it has passed through 
the interferometer, and the interference pattern is recorded with a camera 
focused to image the diffuser.
\label{mich}}
\end{figure}

\clearpage

\begin{figure}
\plottwo{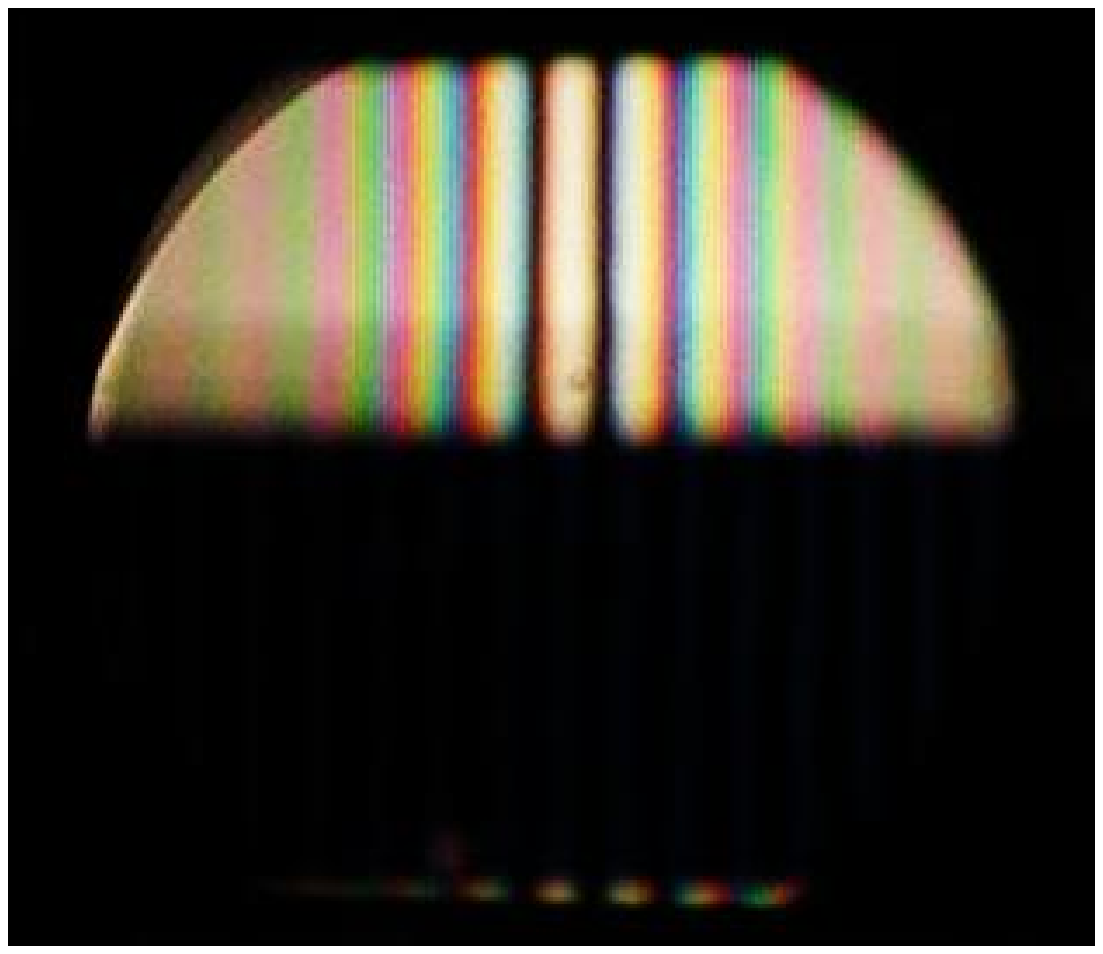}{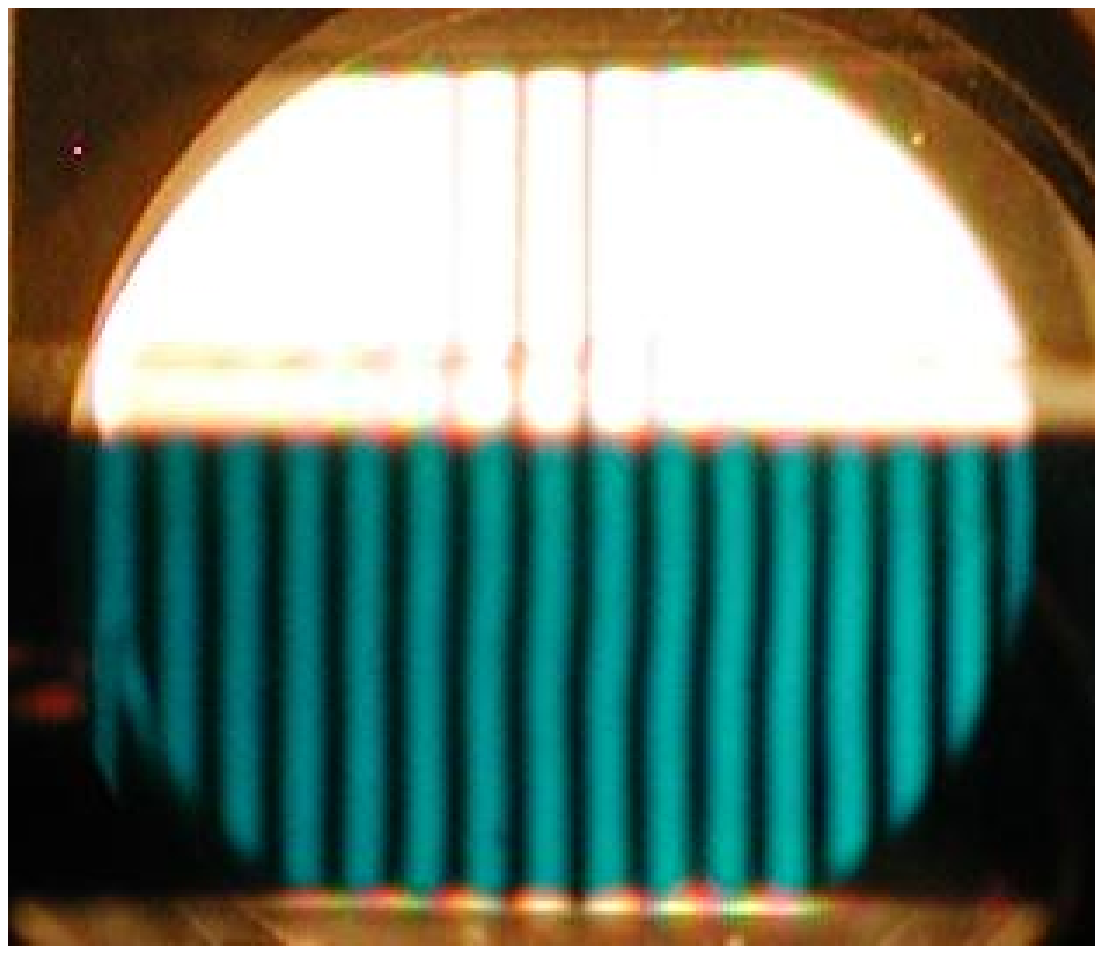}
\caption{{\bf Interference Patterns.}
Both panels show the diffuser photographed through the Michelson interferometer.
The vertical stripes are due to interference. The interferometer is set for
zero path difference along a vertical line through the center of the
image.  One of the interferometer mirrors has been tilted so
the path length difference increases to the right and left of the centerline.
The bottom half of the exit aperture is
covered by a narrow-band filter.
In the left panel the exposure is set so detail in the white light 
interference pattern can be seen in the top half of the image. 
The visibility of the pattern falls nearly to zero in just a few fringes
because the coherence length of the white light is just a few times the
peak wavelength. 
In the right panel the exposure
has been increased so that the narrow band interference pattern, photographed
through the filter, can be seen. Interference visibility is 
maintained across the entire width of the pattern because the filter has
increased the coherence length of the light.  This shows that coherence
can be established in the light waveform after the light has passed
through the interference apparatus. 
\label{fringes}}
\end{figure}

\clearpage

\begin{figure}
\plotone{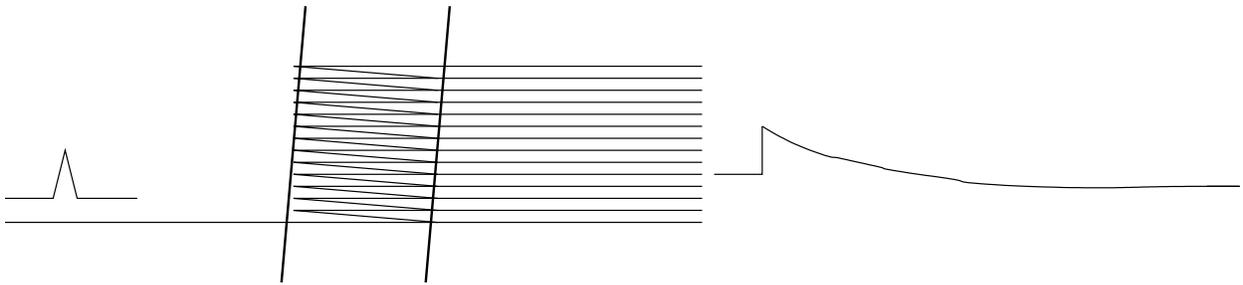}
\caption{{\bf Light Storage in a Bandpass Filter.} 
A narrow passband filter stores light for a long time, 
increasing the coherence length of the stored waves. Here multiple
reflections in a Fabre-Perot Etalon 
are shown.
\label{fabre-perot}}
\end{figure}

\end{document}